\documentclass[preprint,showpacs,preprintnumbers,amsmath,amssymb]{revtex4}
\usepackage{epsfig}
\usepackage{graphicx}
\usepackage{dcolumn}
\usepackage{bm}

\newcommand{\ord}{{\cal O}}
\def\beq{\begin{equation}}
\def\eeq{\end{equation}}
\def\eeqn{\end{equation}}
\newcommand\iden{\leavevmode\hbox{\small1\normalsize\kern-.33em1}}

\newcommand{\sq}{\sqrt{2}}
\newcommand{\bea} {\begin{eqnarray}}
\newcommand{\eea} {\end{eqnarray}}

\newcommand{\Gm}{\Gamma}
\newcommand{\sbt}{s_{\beta}}
\newcommand{\cbt}{c_{\beta}}
\newcommand{\tbt}{t_{\beta}}
\newcommand{\sbcb}{s_{\beta} c_\beta}

\let\jnfont=\rm
\def\NPB#1,{{\jnfont Nucl.\ Phys.\ B }{\bf #1},}
\def\PLB#1,{{\jnfont Phys.\ Lett.\ B }{\bf #1},}
\def\EPJC#1,{{\jnfont Eur.\ Phys.\ Jour.\ C }{\bf #1},}
\def\PRD#1,{{\jnfont Phys.\ Rev.\ D }{\bf #1},}
\def\PRL#1,{{\jnfont Phys.\ Rev.\ Lett.\ }{\bf #1},}
\def\MPLA#1,{{\jnfont Mod.\ Phys.\ Lett.\ A }{\bf #1},}
\def\JPG#1,{{\jnfont J.\ Phys.\ G }{\bf #1},}
\def\CTP#1,{{\jnfont Commun.\ Theor.\ Phys.\ }{\bf #1},}
\def\JHEP#1,{{\jnfont JHEP \ }{\bf #1},}
\def\NPPS#1,{{\jnfont Nucl.\ Phys.\ Proc.\ Suppl.\ }{\bf #1},}
\def\CPC#1,{{\jnfont Computl.\ Phys.\ Commun.\ }{\bf #1},}
\def\CPL#1,{{\jnfont Chin.\ Phys.\ Lett. }{\bf #1},}
\def\APPB#1,{{\jnfont Acta\ Phys.\ Polon.\ B }{\bf #1},}

\def\lsim{\raise0.3ex\hbox{$<$\kern-0.75em\raise-1.1ex\hbox{$\sim$}}}
\def\gsim{\raise0.3ex\hbox{$>$\kern-0.75em\raise-1.1ex\hbox{$\sim$}}}

\begin{document}

\title{\ \\[10mm] Little Higgs theory confronted with
                  the LHC Higgs data}

\author{Xiao-Fang Han$^1$, Lei Wang$^1$, Jin Min Yang$^{2}$,
Jingya Zhu$^2$}

\affiliation{
$^1$ Department of Physics, Yantai University, Yantai 264005, PR China\\
$^2$ State Key Laboratory of Theoretical Physics,\\
     Institute of Theoretical Physics, Academia Sinica,
             Beijing 100190, PR China
\vspace{0.5cm} }


\begin{abstract}
We confront the little Higgs theory with the LHC Higgs search data
(up to 17 fb$^{-1}$ of the combined 7 and 8 TeV run).
Considering some typical models, namely the littlest Higgs model
(LH), the littlest Higgs model with T-parity (LHT-A and LHT-B) and
the simplest little Higgs model (SLH), we scan over the parameter
space in the region allowed by current experiments. We find that in
these models the inclusive and exclusive (via gluon-gluon fusion)
diphoton and $ZZ^*$ signal rates of the Higgs boson are always
suppressed and approach to the SM predictions for a large scale $f$.
Thus, the $ZZ^*$ signal rate is within the $1\sigma$ range of the
experimental data while the inclusive diphoton signal rate is always
outside the $2\sigma$ range. Especially, in the LHT-A the diphoton
signal rate is outside the $3\sigma$ range of the experimental data
for $f < 800$ GeV. We also perform a global $\chi^2$ fit to the
available LHC and Tevatron Higgs data, and find that these models
provide no better global fit to the whole data set (only for some
special channels a better fit can be obtained, especially in the LHT-B).
\end{abstract}


\maketitle

\section{Introduction}
To solve the fine-tuning problem of the standard model (SM), the
little Higgs theory \cite{LH} is proposed as a kind of electroweak
symmetry breaking mechanism accomplished by a naturally light Higgs
sector. So far various realizations of the little Higgs have been
proposed \cite{otherlh,lst,sst}, which can be categorized generally
into two classes \cite{smoking}. One class utilize some product
group, represented by the littlest Higgs model (LH) \cite{lst} in
which the SM $SU(2)_L$ gauge group is from the diagonal breaking of
two (or more) gauge groups. Further, to relax the constraints from
the electroweak precision data \cite{cstrnotparity}, a discrete
symmetry called T-parity is introduced to the LH
\cite{tparity,lhti}. The LH with T-parity (LHT) can provide a
candidate for the cosmic dark matter. The other class use some
simple group, represented by the simplest little Higgs model (SLH)
\cite{sst} in which a single large gauge group is broken down to the
SM $SU(2)_L$. Since these little Higgs models predict different
Higgs property from the SM prediction, they can be tested in the
Higgs search experiments.

Recently, the CMS and Atlas collaborations have announced the
observation of a new boson around 125 GeV \cite{cmsh,atlh}. This
observation is corroborated by the Tevatron search results which
showed a 2.5$\sigma$ excess in the range 115-135 GeV
\cite{12073698:4}. The LHC search results have just been updated by
using 17 fb$^{-1}$ of 7 TeV and 8 TeV data
\cite{17atl,17cms,1211-a-2ph,1211-a-zz,1211-c-zz}. We note that for
the inclusive data, the signal rates of $ZZ^*$ and $WW^*$ are
consistent with the SM values while the diphoton rate is sizably
higher than the SM expectation. For the $Vb\bar{b}$ and $\tau\tau$
channels, the uncertainties are still large.

Although so far the inclusive Higgs search data is roughly consistent
with the SM predictions, the diphoton enhancement has been
explained in various new physics models, such as the SUSY models \cite{susy},
the two-Higgs-doublet models \cite{2hdm}, the Higgs triplet model \cite{triplet},
the models with extra dimensions \cite{extrad} and other extensions of Higgs models
\cite{hrrmodel}.
For the little Higgs theory, the Higgs property (especially the diphoton decay)
was thoroughly studied \cite{lhhrr} even before the LHC Higgs data.
In this work we use the latest LHC Higgs data
 to check the status of the little Higgs theory.
For this purpose, we will examine some typical models, namely the
littlest Higgs model (LH), the littlest Higgs model with T-parity
(LHT-A and LHT-B) and the simplest little Higgs model (SLH). The
model predictions for the Higgs signal rates will be compared with
the experimental data. Also we will perform a global $\chi^2$ fit to
the available LHC and Tevatron Higgs data \cite{1212-Gunion} to
figure out if the little Higgs theory can provide a better fit than
the SM. Also, we will show the Higgs couplings and some
exclusive signal rates in comparison with the Higgs data as well as
the SM predictions.

Our work is organized as follows. In Sec. II we recapitulate the
little Higgs models. In Sec. III we confront the model predictions
for the Higgs signal rates with the experimental data. Finally, we give our
conclusion in Sec. IV.

\section{little Higgs models}

\subsection{Littlest Higgs model (LH)}
The LH model \cite{lst} consists of a nonlinear sigma model with a
global $SU(5)$ symmetry which is broken down to $SO(5)$ by a vacuum
expectation value (VEV) $f$. A subgroup $[SU(2) \otimes U(1)]^2$ of
$SU(5)$ is gauged. The heavy gauge bosons ($W_H$, $Z_H$, $A_H$),
triplet scalar ($\Phi^{++}$, $\Phi^+$, $\Phi^0$, $\Phi^P$) and top
quark partner $T$ quark are respectively introduced to cancel the
Higgs mass one-loop quadratic divergence contributed by the gauge
bosons, Higgs boson and top quark of the SM. There masses are given
as
\begin{align}
        &m_{Z_H} = m_{W_H} = \frac{gf}{2sc},
  \qquad
  m_{A_H} = \frac{g^{\prime}f}{2\sqrt{5}s^{\prime}c^{\prime}},\nonumber\\
        &m_{\Phi}= \frac{\sqrt{2}m_h}{\sqrt{1-x^2}}\frac{f}{v},
        ~~~ \qquad m_T=\frac{m_tf}{s_t c_t v} ,\label{ahvv2}
\end{align}
where $h$ and $v$ are respectively the SM-like Higgs boson and its
vacuum expectation value (vev), $c$, $s\equiv\sqrt{1-c}$, $c'$ and
$s'\equiv\sqrt{1-c'}$ are the mixing parameters in the gauge boson
sector, $x$ is a free parameter of the Higgs sector proportional to
the triplet vev $v'$ and defined as $x = 4fv'/v^2$,
$c_t$ and $s_t\equiv\sqrt{1-c_t}$ are the mixing parameters
between $t$ and $T$.

The relevant Higgs couplings are given as \cite{frlh,hrrhan}
\begin{eqnarray}\label{lhvinter}
 {\cal L}&=&
  2\frac{m_{W_H}^2}{v} y_{_{W_H}} W^{+}_H W^{-}_H h + 2\frac{m_{W}^2}{v} y_{_W} W^+ W^- h
 +2\frac{m_{Z}^2}{v} y_{_Z} Z Z h\nonumber\\
&&
 -2\frac{m_{\Phi}^2}{v} y_{_{\Phi^+}} \Phi^+ \Phi^- h
 - 2\frac{m_{\Phi}^2}{v} y_{_{\Phi^{++}}} \Phi^{++}
  \Phi^{--} h,\nonumber\\
&& - \frac{m_T}{v} y_{_T}\bar{T}T h-\frac{m_t}{v} y_{t} \bar{t}t
h-\frac{m_{\rm{f}}}{v} y_{\rm{f}} \bar{\rm{f}}\rm{f} h
~(\rm{f}=b,~\tau)
 \end{eqnarray}
with
\begin{eqnarray}
&& y_{_{W_H}}=- s^2c^2\frac{v^2}{f^2},\nonumber\\
&& y_{_{W}}=1+
\frac{v^2}{f^2}\left[-\frac{1}{6}-\frac{1}{4}(c^2-s^2)^2\right]=1+
\frac{v^2}{f^2}\left[-\frac{5}{12}+c^2 s^2\right],\nonumber\\
&& y_{_{Z}}=1+
\frac{v^2}{f^2}\left[-\frac{1}{6}-\frac{1}{4}(c^2-s^2)^2
-\frac{5}{4}(c^{\prime 2}-s^{\prime
2})+\frac{1}{4}x^2\right],\nonumber\\
&& y_{\Phi^+}=
\frac{v^2}{f^2}\left[-\frac{1}{3}+\frac{1}{4}x^2\right], \nonumber\\
&& y_{\Phi^{++}}= \frac{v^2}{f^2}\ord(\frac{x^2}{16}\frac{v^2}{f^2},\frac{1}{16\pi^2}), \nonumber\\
&& y_{_{T}} = -c_t^2s_t^2\frac{v^2}{f^2},\nonumber\\
&& y_{t} = 1+\frac{v^2}{f^2} \left[ -
\frac{2}{3}+\frac{x}{2}-\frac{x^2}{4} + c_t^2 s_t^2
\right],\nonumber\\
&& y_{b,~\tau}=1+\frac{v^2}{f^2} \left[ -
\frac{2}{3}+\frac{x}{2}-\frac{x^2}{4}\right]. \label{yxishu}
\end{eqnarray}
In the LH model the relation between
$G_F$ and $v$ is modified from its SM form, which can induce
\cite{hrrhan} \beq v \simeq
v_{SM}[1-\frac{v^2_{SM}}{f^2}(-\frac{5}{24}+\frac{1}{8}x^2)], \eeq
where $v_{SM}=246$ GeV is the SM Higgs vev.

\subsection{Littlest Higgs models with T-parity (LHT)}
T-parity requires that the coupling constant of $SU(2)_1$ ($U(1)_1$)
equals to that of $SU(2)_2$ ($U(1)_2$), which leads to that the four
mixing parameters in gauge sector $c$, $s$, $c'$ and $s'$ equal to
$1/\sqrt{2}$, respectively. Under T-parity, the SM bosons are T-even
and the new bosons are T-odd. Therefore, the coupling $H^{\dag}\phi
H$ is forbidden, leading to the triplet vev $v'=0$ and $x=0$. Since the
correction of $W_H$ to the relation between $G_F$ and $v$ is
forbidden by T-parity, the Higgs vev $v$ is different from that of
the LH \cite{lhtiyuan}, which is
\beq v\simeq
v_{SM}(1+\frac{1}{12}\frac{v^2_{SM}}{f^2}). \eeq Taking
$c=s=c'=s'=1/\sqrt{2}$ and $x=0$, we can obtain the Higgs couplings
to the gauge bosons and scalars of the LHT from
Eq.(\ref{lhvinter}) and Eq. (\ref{yxishu}).

For each SM quark (lepton), a heavy copy of mirror quark (lepton)
with T-odd quantum number is added in order to preserve
T-parity. The Higgs couplings to each generation of mirror quarks are
given by \cite{lhtiyuan}
\begin{eqnarray} {\cal L}_{\kappa}&\simeq& -\sqrt{2} \kappa f
\left[\frac{1+c_\xi}{2} \bar{u}_{L_-} u'_R
-\frac{1-c_\xi}{2}\bar{u}_{L_-}q_R
-\frac{s_\xi}{\sqrt{2}}\bar{u}_{L_-} \chi_R\right] \nonumber \\
&& - m_q \bar{q}_Lq_R - m_{\chi}\bar{\chi}_L\chi_R +{\rm
h.c.}\label{lhti-odd}
\end{eqnarray}
with $c_\xi \equiv \cos\frac{v+h}{\sqrt{2}f}$ and $s_\xi\equiv
\sin\frac{v+h}{\sqrt{2}f}$. After diagonalization of the mass matrix
in Eq. (\ref{lhti-odd}), we can get the T-odd mass eigenstates
$u_-$, $q$ and $\chi$ as well as their couplings to Higgs boson.

For the implementation of T-parity in the Yukawa sector of the
top quark, the T-parity images for the original top quark
interaction of the LH is introduced to make the Lagrangian
T-invariant \cite{lhtiyuan,lhtihubisz},
\beq
{\cal L}_t\simeq
-\lambda_1 f \left[\frac{s_\Sigma}{\sqrt{2}} \bar{u}_{L_+}u_R+
\frac{1+c_\Sigma}{2} \bar{U}_{L_+}u_R \right]-\lambda_2f
\bar{U}_{L_+}U_{R_+}+{\rm h.c.}
\label{lhti-t}
\eeq
with
$c_\Sigma\equiv \cos\frac{\sqrt{2}(v+h)}{f}$ and $s_\Sigma\equiv
\sin\frac{\sqrt{2}(v+h)}{f}$. The mass eigenstates $t$ and $T$ can
be obtained by mixing the interaction eigenstates in Eq.
(\ref{lhti-t}). The mixing parameters are the same to $c_t$ and
$s_t$ of the LH, which are given by
\beq
r=\frac{\lambda_1}{\lambda_2},~~ c_t=\frac{r}{\sqrt{r^2+1}},~~
s_t=\frac{1}{\sqrt{1+r^2}}.
\eeq
The Higgs couplings to $t$ and $T$ are the same to those of LH with $x=0$.

For the SM down-type quarks (leptons), the Higgs couplings of LHT
have two different cases \cite{lhtiyuan}:
\begin{eqnarray}
\frac{C_{hd\bar{d}}}{C_{hd\bar{d}}^{\rm SM}}
&\simeq&1-\frac{1}{4}\frac{v_{SM}^2}{f^2}+\frac{7}{32}
\frac{v_{SM}^4}{f^4} ~~~~{\rm for~LHT-A}, \label{Higgs-downA} \nonumber\\
&\simeq&1-\frac{5}{4}\frac{v_{SM}^2}{f^2}-\frac{17}{32}
  \frac{v_{SM}^4}{f^4} ~~~~{\rm for~LHT-B}.\nonumber
\label{eq15}
\end{eqnarray}
The relation of down-type quark couplings also applies to the lepton
couplings.

\subsection{Simplest little Higgs model (SLH)}
The SLH \cite{sst} model has an $[SU(3) \times U(1)_X]^2$ global
symmetry. The gauge symmetry $SU(3) \times U(1)_X$ is broken down to
the SM electroweak gauge group by two copies of scalar fields
$\Phi_1$ and $\Phi_2$, which are triplets under the $SU(3)$ with
aligned vevs $f_1$ and $f_2$.

The new heavy charged gauge boson $W^{'\pm}$ can contribute to the
effective $h\gamma\gamma$ coupling. The Higgs couplings to $W'W'$,
WW and ZZ are given by \cite{slhvdefine}
\beq
 {\cal L}=
 2\frac{m_{W'}^2}{v} y_{_{W'}} W^{'+} W^{'-} h+2\frac{m_{W}^2}{v} y_{_W} W^+ W^- h+
 2\frac{m_{Z}^2}{v} y_{_Z} Z Z h,
 \eeq
where
\begin{eqnarray}
&& m^2_{W^{'+}} = \frac{g^2}{2}f^2,\\
&& y_{_{W'}} \simeq -\frac{v^2}{2f^2},\\
&& y_{_W} \simeq \frac{v}{v_{SM}} \left[ 1- \frac{v_{SM}^2}{
4f^2}\frac{\tbt^4-\tbt^2+1}{\tbt^2} \right],\\
&& y_{_Z} \simeq \frac{v}{v_{SM}}
\left[ 1- \frac{v_{SM}^2}{ 4f^2}\left(\frac{\tbt^4-\tbt^2+1}{\tbt^2}
+ (1-t_W^2)^2\right) \right]
\end{eqnarray}
with $f=\sqrt{f_1^2+f_2^2}$, $t_\beta\equiv
\tan\beta=f_2/f_1$, $c_\beta=f_1/f$,
$s_\beta=f_2/f$ and $t_W=\tan\theta_W$.

The gauged $SU(3)$ symmetry promotes the SM fermion doublets into
$SU(3)$ triplets. The Higgs interactions with the quarks are given
by \bea \label{tTmixing} {\cal L}_t &\simeq&-f \lambda_2^t \left[
x_\lambda^t c_\beta t_1^{c'}(-s_1t'_L
   +c_1T'_L)+s_\beta t_2^{c'} (s_2 t'_L+ c_2 T'_L)\right]+h.c.,\,\\
   \label{dDmixing}
{\cal L}_{d} &\simeq&-f \lambda_2^{d} \left[ x_\lambda^{d} c_\beta
d_1^{c'}
  (s_1 d'_{L}+c_1 D'_{L})+s_\beta d_2^{c'} (-s_2 d'_{L}+c_2 D'_{L})\right]+h.c.,\,\\
\label{sDmixing} {\cal L}_{s} &\simeq&-f \lambda_2^{s} \left[
x_\lambda^{s} c_\beta s_1^{c'}
  (s_1 s'_{L}+c_1 S'_{L})+s_\beta s_2^{c'} (-s_2 s'_{L}+c_2 S'_{L})\right]+h.c.,\, ,
\eea where \bea s_1\equiv \sin {t_\beta (h+v)\over \sqrt{2}f},\ \
s_2\equiv \sin{(h+v) \over \sqrt{2}t_\beta f}.
 \eea
After diagonalization of the mass matrix in Eqs. (\ref{tTmixing}),
(\ref{dDmixing}) and (\ref{sDmixing}), we can get the mass
eigenstates $(t,~T)$, $(d,~D)$ and $(s,~S)$ as well as their
couplings to Higgs boson.

The Higgs couplings to $b\bar{b}$ and $\tau\tau$ normalized to the
SM values are \beq \frac{C_{hb\bar{b}}}{C_{hb\bar{b}}^{\rm
SM}}=\frac{C_{h\tau\tau}}{C_{h\tau\tau}^{\rm SM}} \simeq
\frac{v_{\rm SM}}{v}\left[1-\frac{1}{6s_\beta^2
c_\beta^2}\frac{v^2}{f^2}\right]. \label{eq15} \eeq The SLH model
predicts a pseudo-scalar $\eta$, which obtains the mass via a
tree-level $\mu$ term, \beq -\mu^2 (\Phi^\dagger_1 \Phi_2 + h.c.) =
- 2 \mu^2 f^2 \sbt\cbt \cos\left( \frac{\eta}{\sq \sbt\cbt f}
\right)
 \cos \left(
 \frac{\sqrt{H^\dagger H}}{f \cbt\sbt}
\right)
\end{equation}
with $H$ being the SM-like Higgs doublet field.

In the SLH, the relation between $G_F$ and $v$ is modified from its
SM form, which can induce \cite{slhvdefine} \beq \label{eq:v} v
\simeq v_{SM} \left[ 1+ \frac{v_{SM}^2}{12
f^2}\frac{\tbt^4-\tbt^2+1}{\tbt^2} - \frac{v_{SM}^4}{180
f^4}\frac{\tbt^8-\tbt^6+\tbt^4-\tbt^2+1}{\tbt^4} \right].
\end{equation}

\section{Higgs properties confronted with the Higgs data}
\subsection{Calculations}
As an effective theory, the Higgs potential of little Higgs models
is affected by the theory at the cutoff scale \cite{uv}. We assume
that there are large direct contributions to the potential from
the physics at the cutoff, so that the constraints of Higgs mass on the
parameter space of the little Higgs models are loosened greatly. In
our calculations, the Higgs mass is fixed as 125.5 GeV. We consider
the relevant QCD and electroweak corrections using the code Hdecay
\cite{hdecay}. For the Higgs productions and decays, the little
Higgs models give the corrections by directly modifying the Higgs
couplings to the relevant SM particles.

For the loop-induced decays $h \to gg$ and $h\to\gamma\gamma$, the
little Higgs models give the partial corrections via the reduced
$ht\bar{t}$ and $hWW$ couplings, respectively. Besides, $h\to gg$
can be corrected by the loops of heavy partner quark $T$ in the LH,
$T$ and T-odd quarks in the LHT, and $T,~D$ and $S$ in the SLH. In
addition to the loops of the heavy quarks involved in the $h\to gg$,
the decay $h\to \gamma\gamma$ can be also corrected by the loops of
$W_H$, $\Phi^+$, $\Phi^{++}$ in the LH and LHT and by $W'$ in the
SLH. Note that the LHT and SLH also predict some neutral heavy
neutrinos, which do not contribute to the $h\gamma\gamma$ coupling
at the one-loop level. Although the charged heavy leptons are
predicted by the LHT, they do not have direct couplings with the
Higgs boson.

In the LH the new
free parameters are $f,~c,~c',~c_t$ and $x$.
We scan over these parameters in the ranges:
\beq
1~{\rm TeV}<f<3.5~{\rm TeV},~~ 0<c<1,~~0<c'<1,~~0.45<c_t<0.9,~~0<x<1.
\eeq
Since the
$h\Phi^{++}\Phi^{--}$ coupling is very small, the contributions of
the doubly-charged scalar to the effective $h\gamma\gamma$ coupling
can be ignored.

In the LHT, the new T-odd quarks can give the additional
contributions to the $h \to gg$ and $h\to\gamma\gamma$ via the
loops, which are not sensitive to the actual values of their masses
as long as they are much larger than half of the Higgs boson mass.
The parameters $c=s=c'=s'=1/\sqrt{2}$ and $x=0$ are fixed by
T-parity. T-parity can relax the constraints of the electroweak
precision data sizably, leading to a scale $f$ as low as 500 GeV
\cite{flht-i}. In our calculation  we scan $f$ in the range of
0.5-3.5 TeV.

In the LH and LHT, the parameter $c_t$ determines the Higgs
couplings to $t$, $T$ and $m_T$, and is involved in the calculations
of the $h\to gg$ and $h\to \gamma\gamma$. The $c_t$ dependence of
the top quark loop and $T$ quark loop can cancel to a large extent
(see Eq. (\ref{yxishu})). For $0.45<c_t<0.9$, the corresponding
parameter $r$ varies from 0.5 to 2.0, which is favored by the
electroweak precision data \cite{flht-i}. Besides, for the LH, the
$c$ and $s$ dependence of $W_H$ loop and $W$ loop in the $h\to
\gamma\gamma$ decay can cancel each other to some extent (see Eq.
(\ref{yxishu})). The parameter $x$ can affect $\sigma(gg\to h)$ and
$\Gamma(h\to b\bar{b})$, but the effects of $x$ on  $\sigma(gg\to
h)$/$\Gamma(h\to b\bar{b})$ are weakened to a large extent (see Eq.
(\ref{yxishu})). For $m_h=125.5$ GeV, the decay $h\to A_H A_H$ is
kinematically forbidden in the LH and LHT.

In the SLH, we take $f, ~t_\beta,~ m_T, ~m_D, ~m_S$ and $m_\eta$ as
new free parameters. Ref. \cite{sst} shows that the LEP-II data
requires $f>2$ TeV. Here, we assume the new flavor mixing matrices
in lepton and quark sectors are diagonal \cite{smoking,lpv} so that
$f$ and $t_\beta$ are free from the experimental constraints of the
lepton and quark flavor violating processes. Besides, the
contributions to the EWPD can be suppressed by a large $t_\beta$
\cite{sst,newslh}. For the perturbation to be valid, $t_{\beta}$
cannot be too large for a fixed $f$.  We require
$\ord(v_0^4/f^4)/\ord(v_0^2/f^2) < 0.1$ in the expansion of $v$. The
small masses of $d$-quark and $s$-quark require that $x_\lambda^{d}$
and $x_\lambda^{s}$ are very small. So there is almost no mixing
between the SM down-type quarks and their heavy partners, and the
results are not sensitive to $m_D$ and $m_S$. In addition to the
SM-like decays, the new decays $h\to \eta\eta$ and $h \to Z\eta$ are
open for an enough light $\eta$, whose partial widths are given by
\bea \label{eq:Gamma:new} \Gm(h \to \eta\eta) &=&
 \frac{{\lambda'}^2}{8\pi}\frac{v^2}{m_h} \sqrt{1-x_\eta},\\
\Gamma( h \to Z \eta) &=& \frac{m_h^3}{32 \pi f^2}
  \left( t_\beta - \frac{1}{t_\beta} \right)^2 \,
  \lambda^{3/2} \left(1, \frac{m_Z^2}{m_h^2}, \frac{m_\eta^2}{m_h^2}
 \right ),
\eea
where $\lambda'= - m_\eta^2/[4f^2 s_\beta^2
c_\beta^2\cos(v/\sqrt{2} f \sbcb)]$, $x_\eta
=4m_\eta^2/m_h^2$ and $\lambda (1,x,y) = (1-x-y)^2 - 4 xy$. The
constraint from the nonobservation in the decay $\Upsilon\to\gamma
\eta$ excludes $\eta$ with a mass below 5-7 GeV \cite{prd82-15}.
So, we scan over the following parameter space:
\begin{eqnarray}
&& 2~{\rm TeV} < f < 6~{\rm TeV},~0.5~{\rm TeV}< m_T <3~{\rm TeV}, \nonumber \\
&& 0.5~{\rm TeV}< m_D~(m_S)<3~{\rm TeV},~1<t_\beta<30,~10~{\rm GeV} < m_{\eta} <500
~{\rm GeV}.
\end{eqnarray}

\begin{figure}[tb]
 \epsfig{file=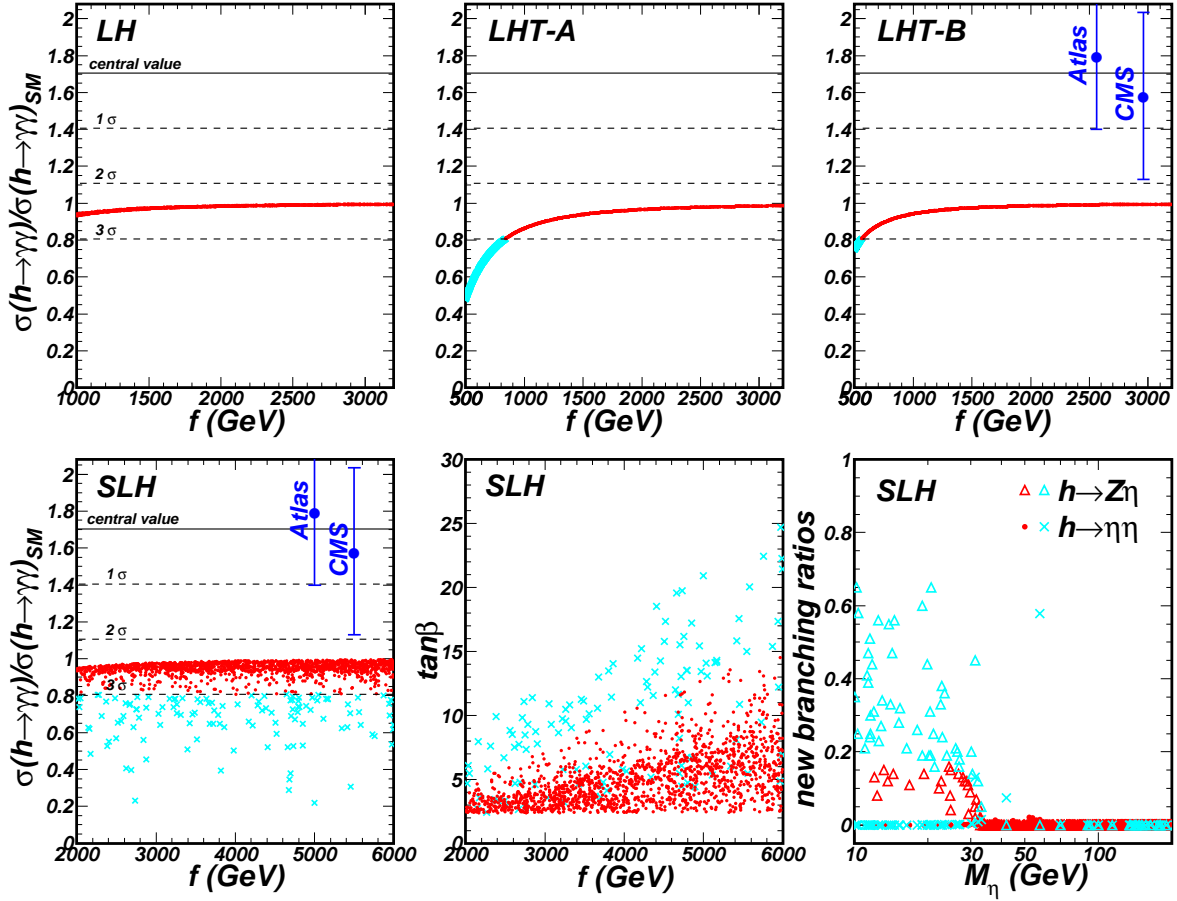,height=12cm}
\vspace{-0.4cm}
\caption{The scatter plots of the parameter space
projected on the planes of the LHC diphoton rate versus $f$,
$\tan\beta$ versus $f$, and the branching ratios of
$h\to Z\eta$ and $h\to \eta\eta$ versus $m_\eta$.
The red and sky-blue samples are respectively within and outside
the $3\sigma$ range of the experimental data of the inclusive diphoton rate.
} \label{fighrr}
\end{figure}

\begin{figure}[tb]
 \epsfig{file=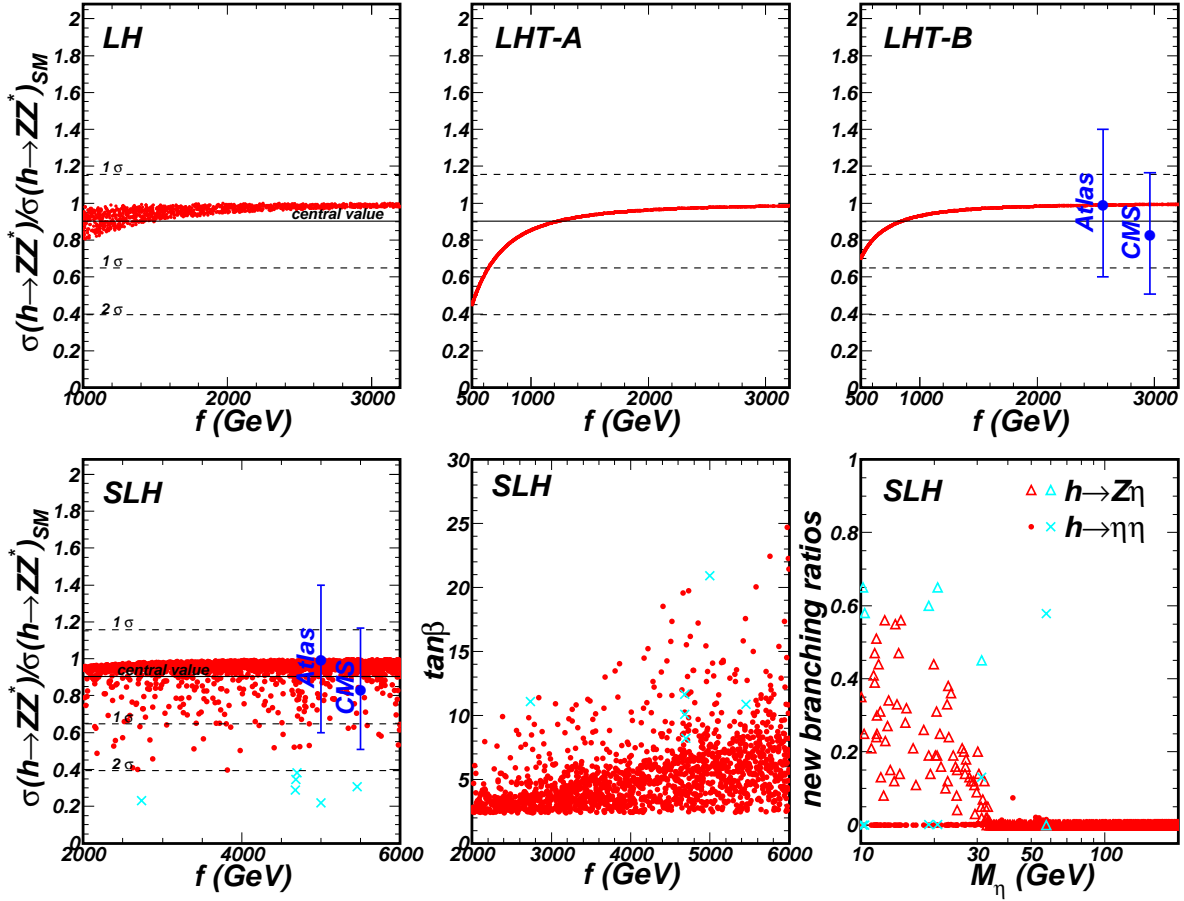,height=12cm}
\vspace{-0.4cm} \caption{Same as Fig. \ref{fighrr}, but showing
the $pp\to h \to ZZ^* \to 4\ell$ signal rate at the LHC.
The red and sky-blue samples are respectively within and outside
the $2\sigma$ range of the experimental data of the inclusive $ZZ^*$ rate.}
\label{fighzz}
\end{figure}

\begin{figure}[tb]
 \epsfig{file=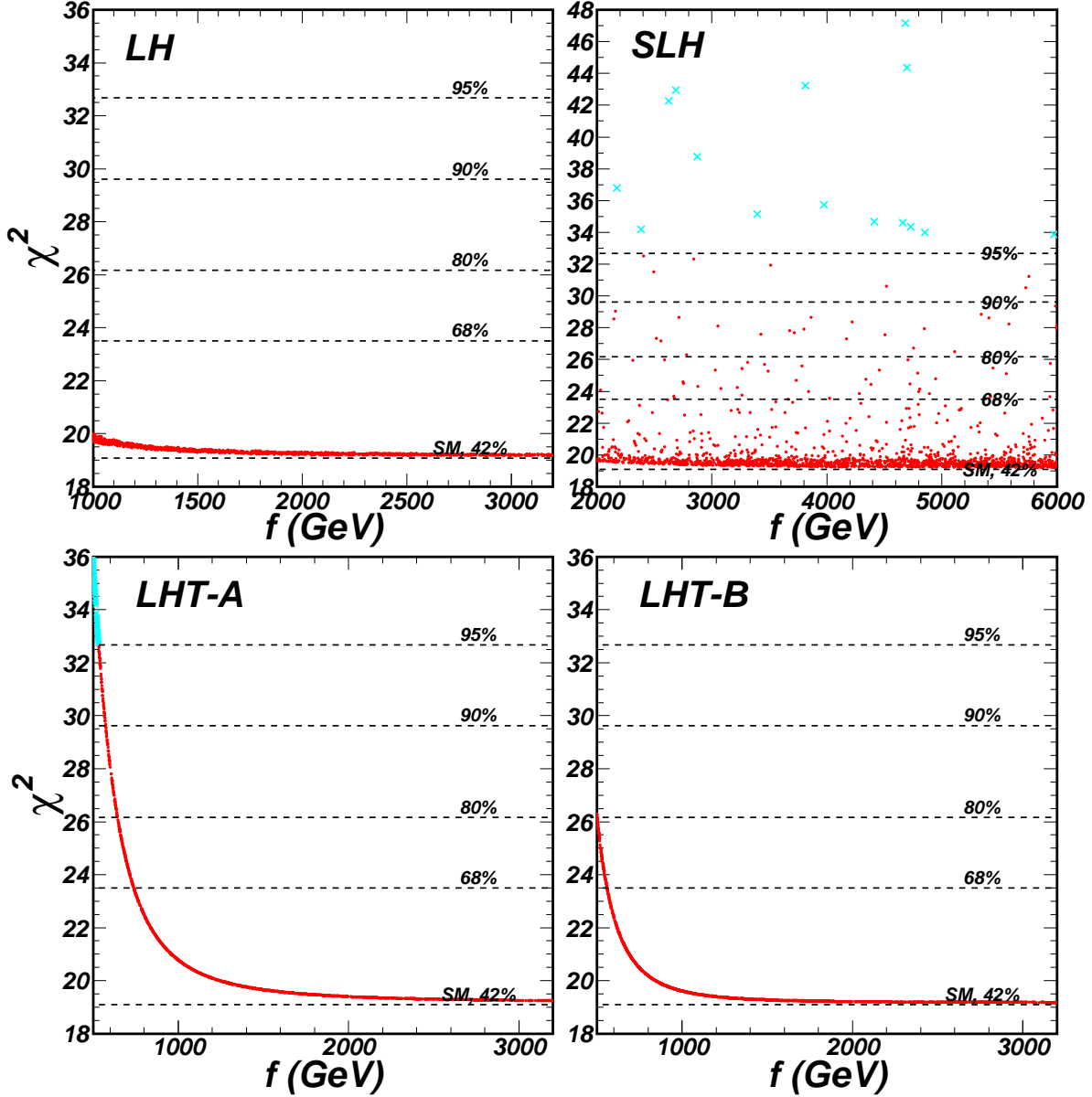,height=16cm}
\vspace{-0.4cm} \caption{The scatter plots of the parameter space
projected on the plane of $\chi^2$ versus $f$. In calculating
$\chi^2$ we use the data of 21 channels \cite{1212-Gunion}.}
\label{figchi}
\end{figure}

\begin{figure}[tb]
 \epsfig{file=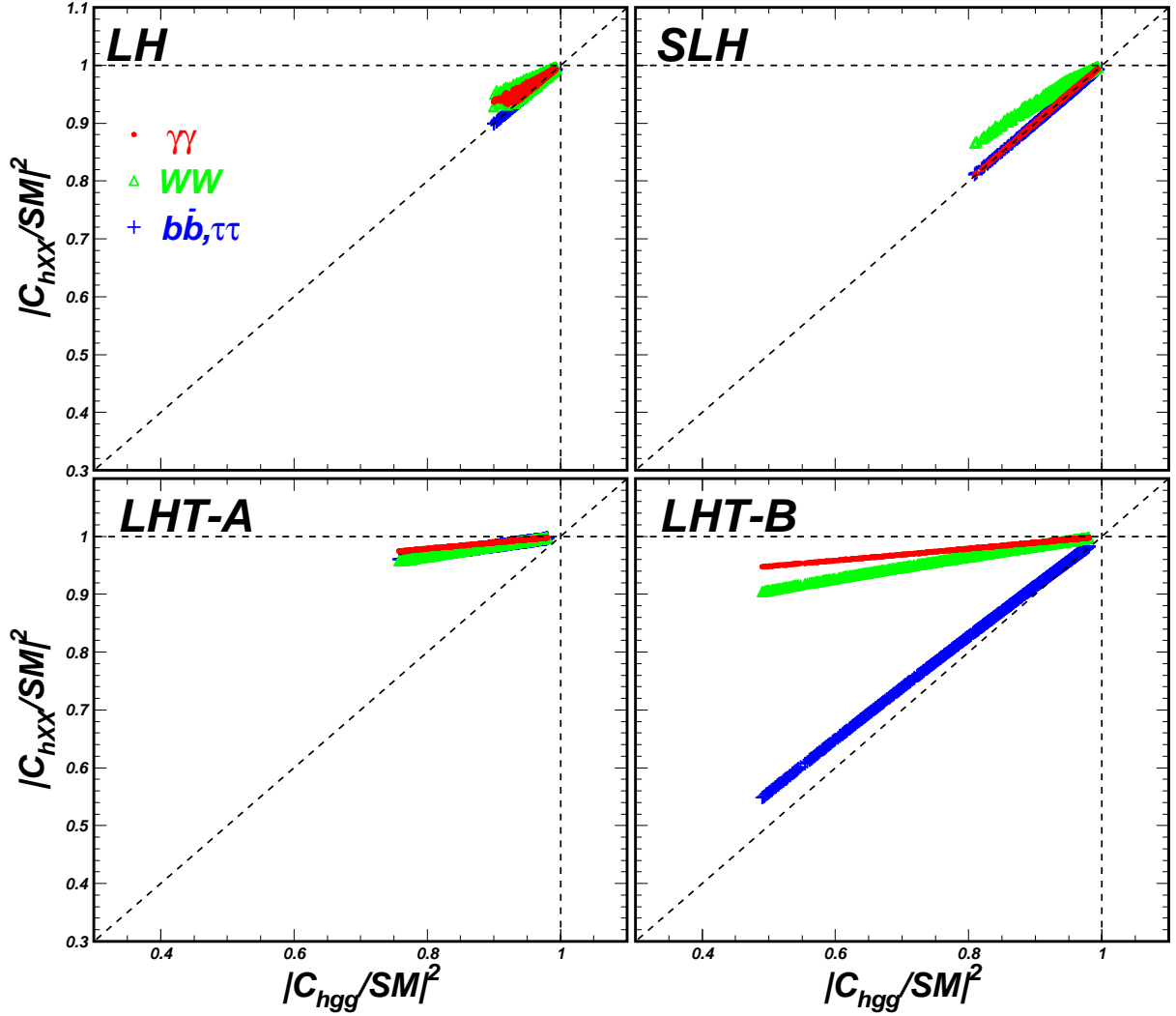,height=14cm}
\vspace{-0.4cm} \caption{
The scatter plots of the parameter space
showing the Higgs couplings normalized to the SM values.
These samples satisfy the conditions: (i) within the $3\sigma$ range of
the diphoton data; (ii) within the $2\sigma$ range of the $ZZ^*$ data;
(iii)  $\chi^2 \leq 32.7$ (corresponding to $95\%$ C.L.).
} \label{figcou1}
\end{figure}

\begin{figure}[tb]
 \epsfig{file=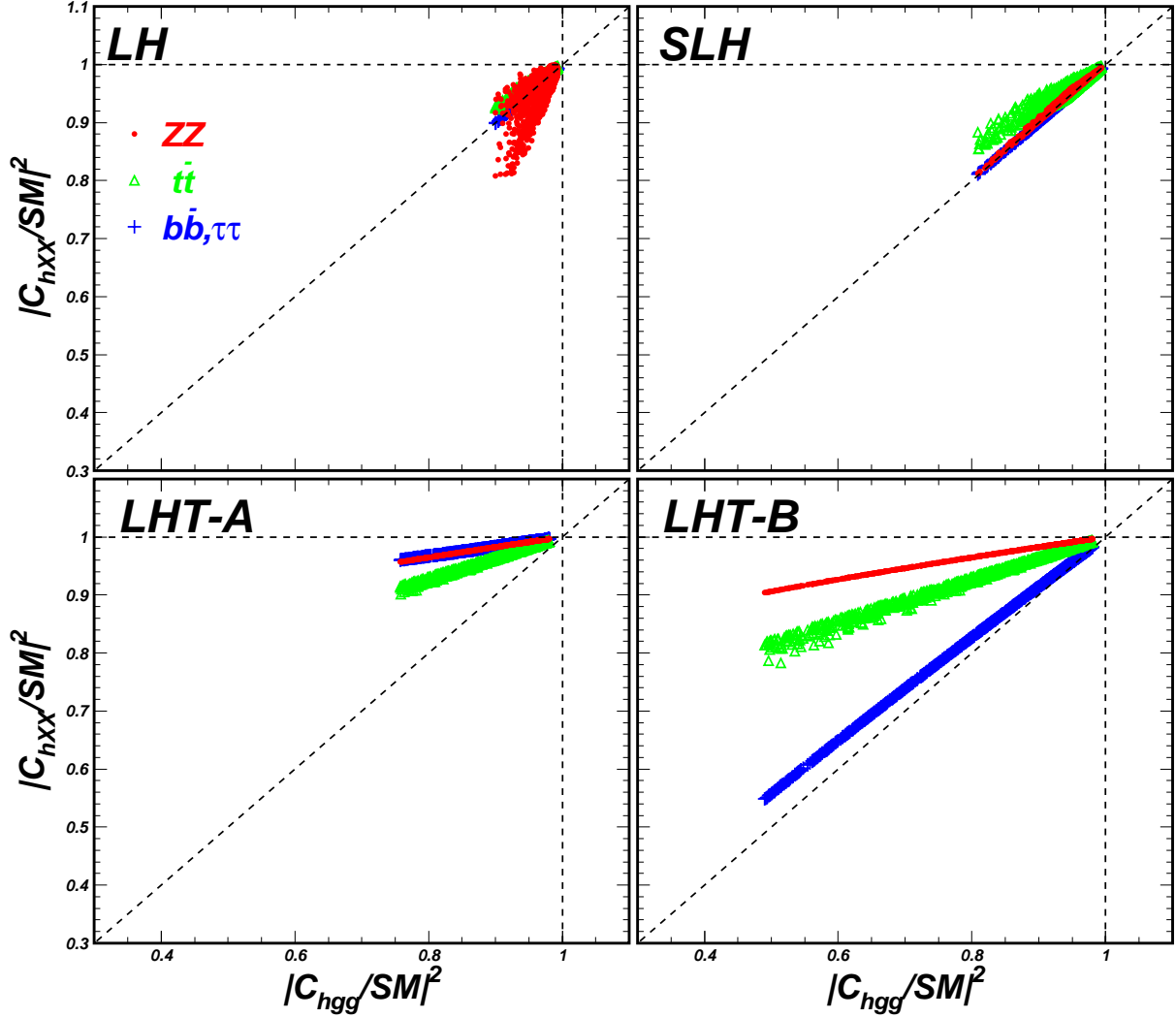,height=14cm}
\vspace{-0.4cm} \caption{Same as Fig. \ref{figcou1}, but showing
different couplings.}
\label{figcou2}
\end{figure}

\begin{figure}[tb]
 \epsfig{file=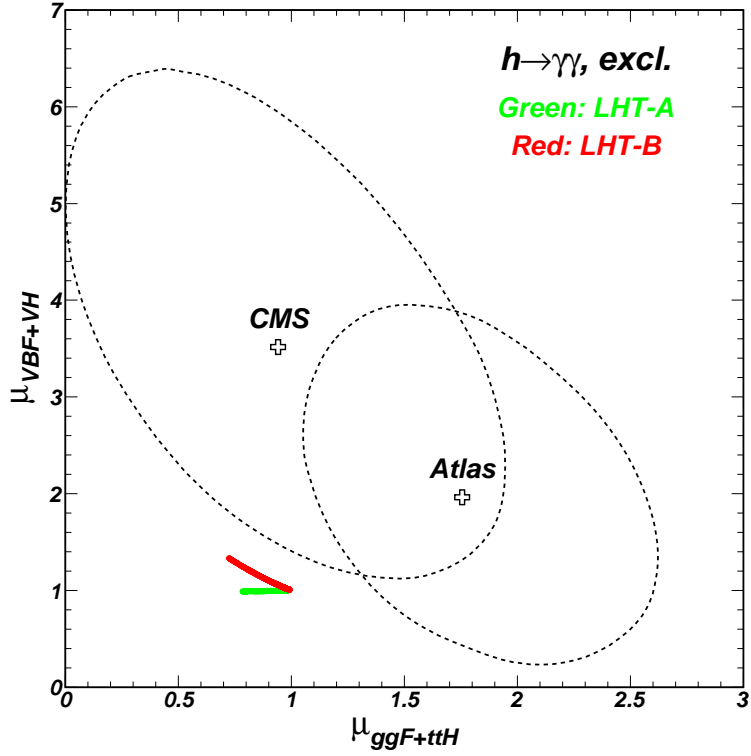,height=10cm}
\vspace{-0.4cm} \caption{The scatter plots of the parameter space
in the LHT-A and LHT-B, showing the exclusive diphoton
rates from the $VBF+VH$ and $ggF+t\bar{t}H$ channels.
The central values and $1\sigma$ contours of the LHC
experiment are taken from \cite{1211-a-2ph,17cms}. } \label{figexcl}
\end{figure}

\begin{figure}[tb]
 \epsfig{file=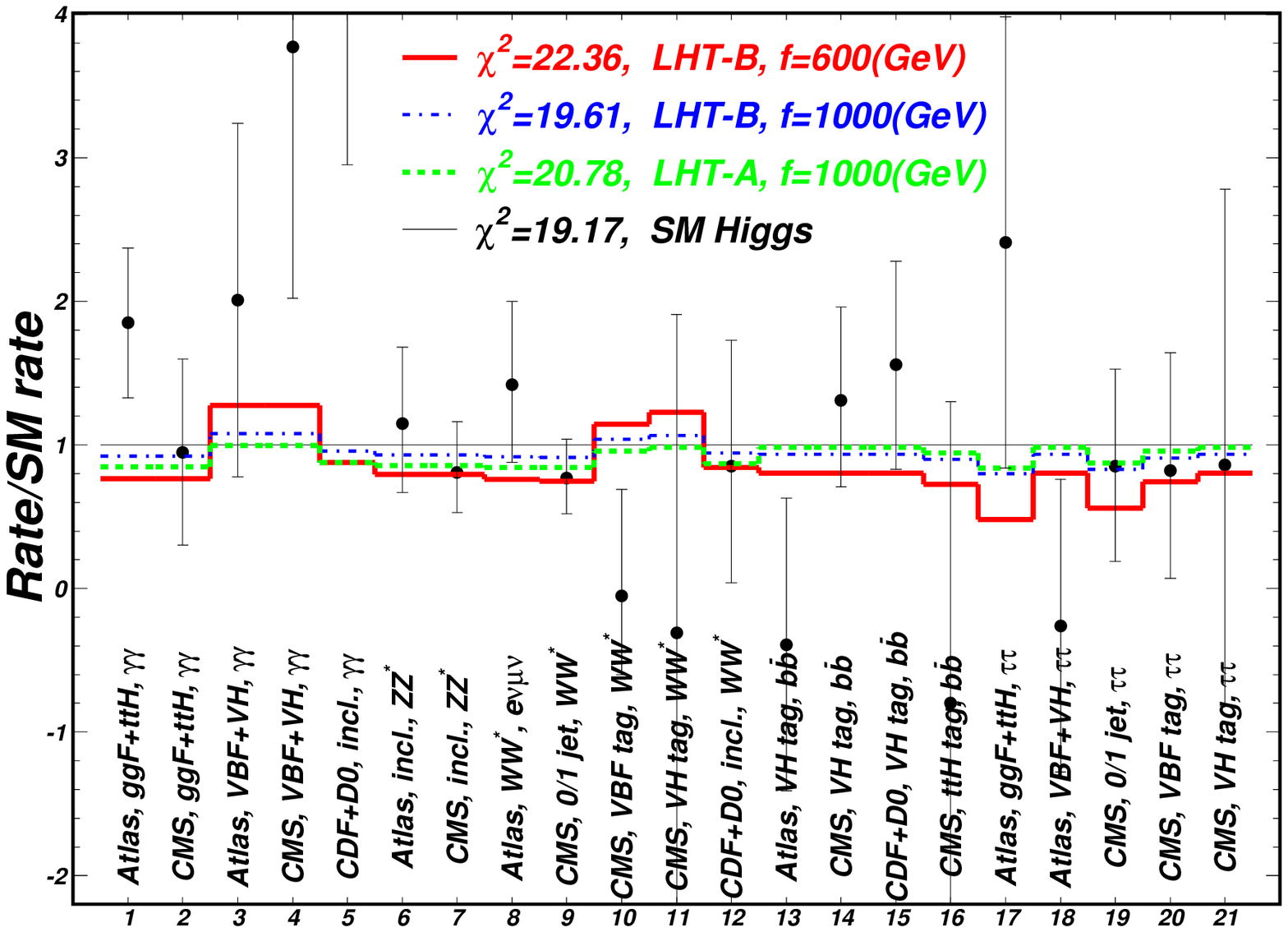,height=11.6cm}
\vspace{-0.4cm} \caption{Predictions of some samples
for various Higgs signal rates at the LHC and Tevatron,
compared with the SM values and the experimental data.
The experiment data is taken from \cite{1212-Gunion}.}
\label{fig3s}
\end{figure}

\subsection{Numerical results and discussions}
The diphoton and $ZZ^*$ are the cleanest channels for the Higgs
boson. We show their inclusive signal rates normalized to the SM
values in Fig. \ref{fighrr} and Fig. \ref{fighzz}, respectively. The
experimental data come from \cite{17atl} for Atlas and \cite{17cms}
for CMS. In combining the data of the two collaborations, we assume
they are independent and Gaussian distributed. Figs. \ref{fighrr}
and \ref{fighzz} show that the rates for the two signals in the
little Higgs models are always suppressed, and approach to the SM
predictions for a large scale $f$.

For the diphoton channel, in these models the signal rates  are
always outside the $2\sigma$ range of the experimental data.
Especially, in the LHT-A the rate is outside the $3\sigma$ range for
$f<800$ GeV. In the SLH the diphoton rate is also sensitive to
$\tan\beta$ and the data favors a small $\tan\beta$. The value of
$m_\eta$ can be as low as 10 GeV, and the total branching ratio of
$h\to Z\eta$ and $h\to \eta\eta$ can only reach 15\% to make the
diphoton rate within the $3\sigma$ range.

For the $ZZ^*$ channel, these models can fit the LHC experimental
data quite well. The signal rate can equal to the central value of
the experimental data for 1 TeV $<f<$ 1.6 TeV in the LH, $f=1.2$ TeV
in the LHT-A, $f=$ 0.8 TeV in the LHT-B, and 2 TeV $<f<$ 6 TeV in
the SLH. For the LH, LHT-A and LHT-B, the rate of $ZZ^*$ is always
within the $2\sigma$ range of the experimental data in the ranges of
parameters scanned. For the SLH, only the parameter space where
 the total branching ratio of $h\to
Z\eta$ and $h\to \eta\eta$ is larger than 60\% are disfavored.

Now we perform a global $\chi^2$ fit to the available LHC and
Tevatron Higgs data in these little Higgs models. We compute the
$\chi^2$ values by the method introduced in \cite{method, hfit-ita}
with the experimental data of 21 channels from \cite{1212-Gunion},
which are shown in Fig. \ref{fig3s}. We assume that the data from
different collaborations or for different inclusive search channels
are independent of each other. However, the data for different
exclusive search channels presented by one collaboration are not
independent, and we use the correlation coefficient given in
\cite{1212-Gunion}. Note that the Higgs mass of $h\to ZZ^* \to
4\ell$ data from Atlas is about 123.5 GeV, which is different from
that of the diphoton data at more than $2\sigma$ level. Therefore,
we rescale the rate at $m_h=123.5$ GeV for a Higgs mass of 125 GeV
and use data
 $\mu(ZZ)=1.15^{+0.53}_{\rm -0.48}$ at $m_h=125$ GeV \cite{1212-Gunion, 1211-a-zz}.
For the case of $\mu(XX)=\hat{\mu}^{+\sigma_+}_{\rm -\sigma_-}$, we
use $\sigma=\sigma_+$ for $\mu(XX)\geq \hat{\mu}$ and $\sigma=\sigma_-$
for $\mu(XX)\leq \hat{\mu}$.

In Fig. \ref{figchi} we project these samples on the plane of
$\chi^2$ versus $f$. We see that the $\chi^2$ values of these models
are larger than the SM value and approach to the SM value for a
sufficiently large $f$ (larger than 2 TeV, 3 TeV, 1.6 TeV and 3 TeV
for the LH, LHT-A, LHT-B and SLH, respectively). Especially, in the
LHT-A the value of $\chi^2$ is larger than 32.7 for $f<$ 530 GeV,
which implies that $f<$ 530 GeV is excluded at 95\% confidence level
from an experimental viewpoint.

Figs. \ref{figcou1} and \ref{figcou2} show the Higgs couplings
normalized to the SM values. We see that in these little Higgs
models the Higgs couplings are all suppressed, and approach to the
SM values for a large $f$. The correlations between the couplings
are also interesting and may be useful for distinguishing different
models. For example, the value of $|C_{hgg}/SM|/|C_{hb\bar{b}}/SM|$
is around 1 for the LH and SLH, but smaller than 1 for the LHT-A and
LHT-B. In the LHT-A and LHT-B, the T-odd quarks further suppress the
$hgg$ coupling, and the suppression is equal compared with that of
top quark and $T$ quark. Note that the reduced $hb\bar{b}$ coupling
can suppress the total width of the 125.5 GeV Higgs boson, which
helps to enhance the branching ratios of $h\to
\gamma\gamma,~WW^*,~ZZ^*, ~\tau\tau$. However, the reduced $hgg$
coupling suppresses the cross section of $gg\to h$ more sizably and
the reduced couplings $h\gamma\gamma$, $hWW$, $hZZ$ and $h\tau\tau$
suppress the width of $h\to\gamma\gamma, WW^*, ZZ^*, \tau\tau$.
Besides, the total width of the Higgs boson in the SLH is enhanced
by the new decay modes $h\to Z\eta$ and $h\to \eta\eta$ for a light
$\eta$ and thus the signal rates are reduced further.

Figs. \ref{figcou1} and \ref{figcou2} show that the Higgs couplings of
LHT-B can be very different from those of the SM, which can lead to
some interesting Higgs phenomena at the colliders. Therefore, we
will pay special attention to this model in the following discusions,
and the LHT-A is also comparatively considered.

\begin{table}
\vspace{-1.5cm} \caption{The detailed information of some samples in
the LHT-A and LHT-B.}
  \setlength{\tabcolsep}{2pt}
  \centering
  \begin{tabular}{|c|c|c|c|}
    \hline \hline
     &~~~~  LHT-B P1 ~~~~& ~~~~LHT-B P2~~~~ &~~~~ LHT-A P3~~~~  \\
    \hline
     $f (GeV)$
     & 601.22 & 999.96&  999.96
     \\
     r & 1.9326 & 1.9286 &  1.9286
     \\
     $\chi^{2}$
     & 22.36  & 19.61  &  20.78
     \\
     $M_T$
     & 1044.27 & 1722.58 &  1722.58
     \\
     $M_{W_H}$
     & 370.47 & 624.92 &  624.92
     \\
     $M_{\Phi}$
     & 427.41 & 717.19 & 717.19
     \\
     $M_{A_H}$
     & 82.85 & 148.54 &  148.54
     \\
     \hline
     $|C_{hgg}/SM|^2$
     & 0.5475 & 0.8265&  0.8265
     \\
     $|C_{hbb}/SM|^2$
     & 0.6013 & 0.8505&  0.9715
     \\
     $|C_{h\tau\tau}/SM|^2$
     & 0.6013 & 0.8505 & 0.9715
     \\
     $|C_{h\gamma\gamma}/SM|^2$
     & 0.9531 & 0.9817& 0.9817
     \\
     $|C_{hWW}/SM|^2$
     & 0.9162 & 0.9697 & 0.9697
     \\
     $|C_{hZZ}/SM|^2$
     & 0.9162 & 0.9697& 0.9697
     \\
     $|C_{htt}/SM|^2$
     & 0.8255 & 0.9322& 0.9322
     \\
     \hline
     LHC, ggF+ttH, $\gamma\gamma$
     & 0.763 & 0.921 & 0.847
     \\
     LHC, VBF+VH, $\gamma\gamma$
     & 1.278 & 1.080 & 0.994
     \\
     Tev, incl., $\gamma\gamma$
     & 0.877 & 0.956 & 0.880
     \\
     LHC, incl., $ZZ^*$
     & 0.797 & 0.929 & 0.855
     \\
     LHC, $WW^*$, $e\nu\mu\nu$
     & 0.759 & 0.917 & 0.844
     \\
     LHC, 0/1 jet, $WW^*$
     & 0.749 & 0.914 & 0.841
     \\
     LHC, VBF tag, $WW^*$
     & 1.144 & 1.040 & 0.957
     \\
     LHC, VH tag, $WW^*$
     & 1.228 & 1.067 & 0.982
     \\
     Tev, incl., $WW^*$
     & 0.843 & 0.944 & 0.869
     \\
      LHC,VH tag, $b\bar{b}$
      & 0.806 & 0.936 & 0.984
     \\
      LHC, ttH tag, $b\bar{b}$
      & 0.726 & 0.900 & 0.946
     \\
      Tev, VH tag, $b\bar{b}$
      & 0.806 & 0.936 & 0.984
     \\
      LHC, ggF, $\tau\tau$
      & 0.482 & 0.798 & 0.839
     \\
      LHC, VBF+VH, $\tau\tau$
      & 0.806 & 0.936 & 0.984
     \\
      LHC, 0/1 jet,  $\tau\tau$
      & 0.559 & 0.831 & 0.873
     \\
      LHC, VBF tag, $\tau\tau$
      & 0.744 & 0.910 & 0.956
     \\
      LHC, VH tag, $\tau\tau$
      & 0.806 & 0.936 & 0.984
     \\
  \hline \hline
  \end{tabular}
\label{chi}\vspace{-0.35cm}
\end{table}

In Fig. \ref{figexcl} we show the exclusive diphoton signal rates
from the $VBF+VH$ and $ggF+ttH$ channels. We can see that, although
the Higgs couplings and inclusive diphoton rate are always reduced,
the exclusive rate of $VBF+VH$ can be enhanced in the LHT-B. The
reason is that the $hb\bar{b}$ coupling in the LHT-B is suppressed
much sizably, which greatly enhances the branching ratio of $h\to
\gamma\gamma$. Therefore, the LHT-B is favored by the enhanced
exclusive diphoton data of $VBF+VH$ from Atlas and CMS (note that
the data has a rather large uncertainty).

Now we take some benchmark points in the LHT-A and LHT-B
to demonstrate the Higgs properties in Table \ref{chi} and
Fig.\ref{fig3s}.
We see that in the LHT-A all the signal rates are
suppressed while in the LHT-B the exclusive signal rates
(except $h \to b\bar{b}$ and $h \to \tau\bar{\tau}$)
of $VBF + VH$ are enhanced, especially for a small $f$.
Compared to the experimental data
shown in Fig. \ref{fig3s}, we find that the LHT-B can provide a
better fit than the SM for some channels like
$VBF  + VH, \gamma\gamma$
of Atlas and CMS, $incl. \ ZZ^*$, $0/ 1 \ jet, WW^*$,
$\tau\tau$ data of CMS.

\section{Conclusion}
In this paper we compared the properties of the SM-like Higgs boson
predicted by the typical little Higgs models (namely the LH, LHT-A, LHT-B and SLH)
with the latest LHC Higgs search data. For a SM-like Higgs boson
around 125.5 GeV, we obtained the following observations:
(i) In these models the inclusive diphoton signal rates cannot
 be enhanced and lie outside the $2\sigma$ range of the present
 data;
(ii) While most signal rates are suppressed in these models,
some exclusive signal rates in the $VBF$ and $VH$ channels
can be enhanced in the LHT-B;
(iii) Compared with the SM, these models provide no better global fit
 to the whole data set, but for some special channels a better fit
can be obtained, especially in the LHT-B;
(iv) In these models the Higgs couplings are
suppressed and approach to the SM values for a sufficiently large
scale $f$.

We should stress again that in little Higgs models the inclusive
diphoton rate cannot get enhanced for obvious reasons. In these
models the $T$-quark (top quark partner) and new heavy gauge bosons
are responsible for canceling the one-loop quadratic divergence of
the Higgs mass contributed by top quark and SM gauge bosons,
respectively. As a result, the Higgs couplings to top quark and
$T$-quark have opposite sign, and the contributions of the $T$-quark
loop will reduce the effective $hgg$ coupling. Similarly, the Higgs
couplings to $W$ boson and the heavy charged gauge boson have
opposite sign, and the contributions of the heavy charged gauge
boson loop will reduce the effective $h\gamma\gamma$ coupling. In
addition, with the expansion of the nonlinear sigma field, the Higgs
couplings to top quark and $W$ boson are suppressed, which will
further reduce the effective $hgg$ and $h\gamma\gamma$ couplings.
However, the $hb\bar{b}$ coupling is also reduced sizably in the LH,
SLH and LHT-B, and thus the signal rates in some channels are quite
close to the SM values.

The future LHC Higgs data (especially the diphoton rate) with large
statistics will allow for a critical test for these little Higgs models.
If the enhancement of the diphoton rates persists
or get enlarged, the little Higgs models will be strongly disfavored or
excluded. Otherwise, if these rates drop below the SM
value, these models will be favored. Also, these models have other
correlated phenomenology like the enhanced Higgs pair production
\cite{hh-lht} and the suppressed $ht\bar{t}$ production
\cite{htt-lht} at the LHC. All these phenomenology can be jointly
utilized to test the little Higgs models and distinguish them from
other new physics models. \vspace{.5cm}

{\em Note added:}
When this manuscript is being prepared, a paper with similar analysis and
similar results appeared in the arXiv \cite{12125930}.

\section*{Acknowledgment}
This work was supported by the National Natural Science Foundation
of China (NNSFC) under grant Nos. 11105116, 11005089, 11275245,
10821504, 11135003 and 11175151.

\end{document}